\documentclass[MSNbibl,nameyear,dvips]{arxstspdf}
\usepackage{flushend}
\usepackage{stfloats}
\volume{29}
\issue{1}
\pubyear{2014}
\firstpage{97}
\lastpage{97}
\doi{10.1214/13-STS459} % kopijuoti is 'New paper accepted'
\begin{document}
\begin{frontmatter}

\title{Contribution by M. A. Girolami}%\thanksref{T1}
% kai straipsnis turi susijusiu diskusiju ir rejoinder'iu
%rejoinder at \relateddoi{r}{10.1214/00-STSXXXX}.}
\runtitle{Contribution by M. A. Girolami}

\begin{aug}
\author[a]{\fnms{Mark A.} \snm{Girolami}\corref{}\ead[label=e1]{m.girolami@ucl.ac.uk}}
\runauthor{M. A. Girolami}

\affiliation{University of Warwick}

\address[a]{Mark A. Girolami is Professor, EPSRC Established Career Research
Fellow, Chair of Statistics, Department of Statistics, University of
Warwick, Coventry, CV4 7AL, United Kingdom \printead{e1}.}

\end{aug}

% ABSTRACT

% KEYWORDS

\end{frontmatter}

This collection of Big Bayes Stories could be partitioned into two groups,
one relating to the sciences, cosmology in particular, and the other
relating to public policy, that is, health, fisheries management and
demographics.

My first comment here is that inferential issues related to the sciences
and the shaping and guiding of public policies can only be addressed
appropriately by adoption of the Bayesian framework. This is a very strong,
and no doubt provocative, statement, the opinion of which has been formed
by my own experience of working very closely with a range of basic
scientists, clinical professionals and econometricians providing
support in
developing fiscal policy.

The almost wholesale adoption of the Bayesian framework by astronomers and
cosmologists is a good case in point where subjective Bayesian
inference is
viewed as a formal codification of the scientific method and therefore most
natural in guiding scientific inquiry.

I have had first-hand experience of this when working with cellular
biologists who previously had viewed statistical analysis as the means of
providing nothing more than the $p$-values required by the editors of
journals such as \textit{Nature}. However, when presented with
the Bayesian
formalism of an expert informed prior to posterior belief updating the\raggedend\vadjust{\vfill\eject}
paradigm has been embraced wholeheartedly by cellular biologists and forms%
the common language of scientific collaboration, for example, Xu et al.
(\citeyear{Xuetal10}).

My second comment is that many of the cases presented required a complex
statistical model, which of course brings with it associated technical
issues, but are most naturally accommodated in the Bayesian framework. When
considering the issues of systematically integrating diverse data sources,
exploiting model structure to employ sparse measurements, or formally and
explicitly quantifying uncertainty induced due to the use of complex
computer codes, it is hard to see how satisfactory and transparent
non-Bayesian solutions would follow.

In summary, I have viewed these interesting case studies from the
perspective of how feasible the required analysis would be as part of an
ongoing dialogue between statisticians and scientists or statisticians and
policy makers. All of them suggest to me that, to misquote Karl Pearson,
Bayesian inference provides the ``Grammar of Science.''


%suskaldyti doi

\renewcommand\bibname{Reference}
% imsref loaded by audrone.aklyte, 2014-02-14 13:30:32
%

\end{document}